\documentclass[
    ,final            
  ,numberedheadings 
  ]
  {aipproc}
\usepackage{graphicx,amssymb,amsmath,times}
\usepackage{subfigure}
\usepackage[dvips]{epsfig}
\usepackage[dvips]{graphicx}
\usepackage[dvips]{color}
\usepackage{epsfig,graphicx}
\usepackage[dvips]{graphicx,color}

\layoutstyle{6x9}

\begin{document}

\title{A New Search Paradigm for Correlated Neutrino Emission from Discrete GRBs using Antarctic Cherenkov Telescopes in the Swift Era\footnote{Contributed to the Proceedings of The 16$^{\text{th}}$ Annual Astrophysics Conference in Maryland: \emph{Gamma Ray Bursts in the Swift Era}. Edited by Stephen S. Holt, Neil Gehrels and John A. Nousek (2006).}}



\author{Michael Stamatikos for the IceCube\footnote{For a full collaboration author list see http://www.icecube.wisc.edu/pub\_and\_doc/conferences/panic05.} $\:$ Collaboration}{
  address={Department of Physics, University of Wisconsin, Madison, WI 53706, USA\footnote{Correspondence to michael.stamatikos@icecube.wisc.edu.}}
}

\author{David L. Band}{
  address={NASA/GSFC, Greenbelt, MD 20771, USA and JCA/UMBC, Baltimore, MD 21250, USA\footnote{Correspondence to dband@milkyway.gsfc.nasa.gov.}}
}

\begin{abstract}
We describe the theoretical modeling and analysis techniques associated with a preliminary search for correlated neutrino emission from GRB980703a, which triggered the Burst and Transient Source Experiment (BATSE GRB trigger 6891), using archived data from the Antarctic Muon and Neutrino Detector Array (AMANDA-B10). Under the assumption of associated hadronic acceleration, the expected observed neutrino energy flux is directly derived, based upon confronting the fireball phenomenology with the discrete set of observed electromagnetic parameters of GRB980703a, gleaned from ground-based and satellite observations, for four models, corrected for oscillations. Models 1 and 2, based upon spectral analysis featuring a prompt photon energy fit to the Band function, utilize an observed spectroscopic redshift, for isotropic and anisotropic emission geometry, respectively. Model 3 is based upon averaged burst parameters, assuming isotropic emission. Model 4, based upon a Band fit, features an estimated redshift from the lag-luminosity relation with isotropic emission. Consistent with our AMANDA-II analysis of GRB030329, which resulted in a flux upper limit of $\sim0.150\:\text{GeV}/\text{cm$^{2}/$s}$ for model 1, we find differences in excess of an order of magnitude in the response of AMANDA-B10, among the various models for GRB980703a. Implications for future searches in the era of Swift and IceCube are discussed.
\end{abstract}

\maketitle


\section{GRB030329 \& The Case for a New Paradigm}

Canonical fireball phenomenology, in the context of hadronic acceleration, predicts correlated MeV to EeV neutrinos from gamma-ray bursts (GRBs). Ideal for detection are $\sim$ TeV-PeV muon neutrinos, which arise as the leptonic decay products of photomeson interactions ($p+\gamma \rightarrow \Delta ^{+} \rightarrow \pi ^{+} + [n] \rightarrow \mu^{+}+\nu_{\mu} \rightarrow e^{+} + \nu_e + \bar{\nu}_{\mu}+\nu_{\mu}$) within the internal shocks of the relativistic fireball. Since the prompt $\gamma$-rays act as the ambient photon target field, these neutrinos are expected to be in spatial and temporal coincidence, with inverted\footnote{Consequently, $\epsilon_{\nu}\propto\epsilon_{\gamma}^{-1}\propto\epsilon_{p}$, as regulated by the normalization ($A_{\nu_{\mu}}$) and proton energy efficiency ($f_{\pi}$).} energy spectra (see Equation~\ref{neutrino_energy_spectrum}), which trace the photon energy spectra (see Equation~\ref{Band_function2}), due to the intrinsic threshold requirement that $4(1+z)^{2}\epsilon_{p}\epsilon_{\gamma}\geq\left(m_{\Delta^{+}}^{2}-m_{p}^{2}\right)\Gamma_{\text{bulk}}^{2}$. Constraints imposed by coincidence are tantamount to nearly background-free searches in neutrino observatories such as AMANDA and IceCube (see Table~\ref{GRB030329_results_summary}). The former, which has been calibrated with atmospheric neutrinos, has demonstrated the viability of high energy neutrino astronomy using the ice at the geographic South Pole as a Cherenkov medium since 1997. The latter, AMANDA's km-scale successor, is currently under construction with anticipated completion by 2010.

A positive detection of such high energy neutrinos would confirm hadronic acceleration in the relativistic GRB-wind, providing critical insight to the associated micro-physics of the fireball, while possibly revealing an astrophysical acceleration mechanism for the highest energy cosmic rays. AMANDA analyses have used a diffuse formulation \cite{WaxmanBahcall:1999}, predicated on an ensemble of average GRBs, to produce the most stringent upper limits upon correlated multi-flavored neutrino emission \cite{Achterberg:2005}. However, over 30 years of ground-based and satellite electromagnetic observations have documented the following facts: (i) the electromagnetic parameters of GRBs are characterized by distributions, often spanning multiple orders of magnitude, which differ both among and within bursts of different classes (i.e. short, long, x-ray rich, etc.), whose deviation from averaged values is often not accommodated by the inherent uncertainty of measurement; (ii) GRB satellite detectors exhibit a large dynamic range of thresholds and sensitivities which impart statistical sample bias via selection effects. These facts render the notion of an \emph{average} GRB incompatible with the observational record. Furthermore, it has been argued that electromagnetic variations may lead to variations in the number of expected neutrino events associated with GRBs \cite{Guettaref:2004}. This reasoning has led to a new modeling paradigm for correlated neutrino emission searches, which is based upon the notion of a \emph{discrete} GRB, i.e. one that is described by a unique set of electromagnetic parameters.

The quantitative effects on the expected neutrino number, energy and subsequent constraints upon astrophysical models (due to null detection), based upon multiple models of emission geometry and electromagnetic characterization, were initially illustrated in an analysis of GRB030329 with AMANDA-II \cite{Stamatikos:2005d}. For GRB030329, a peak effective area for muon neutrinos of $\sim80\:\text{m}^{2}$ and $\sim700\:\text{m}^{2}$ at $\sim2$ PeV and an effective area for muons of $\sim100,000\:\text{m}^{2}$ and $\sim1\:\text{km}^{2}$ at $\sim200$ TeV were achieved for AMANDA-II and predicted for IceCube, respectively. Principal results, including neutrino flux upper limits for each model tested, are summarized in Table~\ref{GRB030329_results_summary}. Further details regarding this analysis may be found elsewhere \cite{Stamatikos:2005d}. Here, supplementary to our conference presentation, we extend the paradigm by taking a first look at GRB980703a, one of the BATSE bursts currently under investigation using AMANDA archived data from 1997-2000 \cite{Stamatikos:2004b,Stamatikos:2006}, using models 1 (discrete-isotropic), 2 (discrete-jet), 3 (average-isotropic) and 4 (lag-isotropic).  Model 4 demonstrates that the lag-luminosity relation \citep{Bandref:2004} provides a reasonable redshift estimate.

\begin{table} [h]
\begin{tabular}{ccccc}
\hline
\small{\textbf{Model}} & \small{$n_{b}$,\:$n_{b}^{\prime}$\tablenote{Number of background events expected during a 40 second on-time search window before ($n_{b}$) and after ($n_{b}^{\prime}$) quality selection (optimized for discovery \citep{Hill:2005}), including restriction to a search bin radius (
space angle between the reconstructed muon trajectory and the GRB's position) of 11.3$^{\circ}$.}} & \small{$N_{s},\:n_{s},\:n_{s}^{\prime}$\tablenote{Number of neutrino signal events expected on-time for IceCube ($N_{s}$) and AMANDA-II ($n_{s}$, $n_{s}^{\prime}$).}} & \small{$n_{obs},\:n_{obs}^{\prime}$\tablenote{The number of observed events in AMANDA-II before ($n_{obs}$) and after ($n_{obs}^{\prime}$) quality selection.}} & \small{Flux Upper Limit\tablenote{Based upon null detection in AMANDA-II. The effects of neutrino flavor oscillations have been included. For more details on GRB030329's electromagnetic and neutrino parameterization, see \cite{Stamatikos:2005d}.} $\frac{\text{GeV}}{\text{cm$^{2}\cdot$s}}$}\\
\hline
1 & 17.44, 0.06 & 0.1308, 0.0202, 0.0156 & 15, 0 & 0.150 \\
2 & 17.44, 0.06 & 0.0691, 0.0116, 0.0092 & 15, 0 & 0.039 \\
3 & 17.44, 0.06 & 0.0038, 0.0008, 0.0006 & 15, 0 & 0.035 \\
\hline
\end{tabular}
\caption{Summary of results for GRB030329 \cite{Stamatikos:2005d}.}
\label{GRB030329_results_summary}
\end{table}

\section{GRB980703a: Extending the Discrete GRB Paradigm}

\begin{figure}[t]
\centering
\begin{minipage}[c]{0.46\textwidth}
\includegraphics[width=1.00\textwidth]{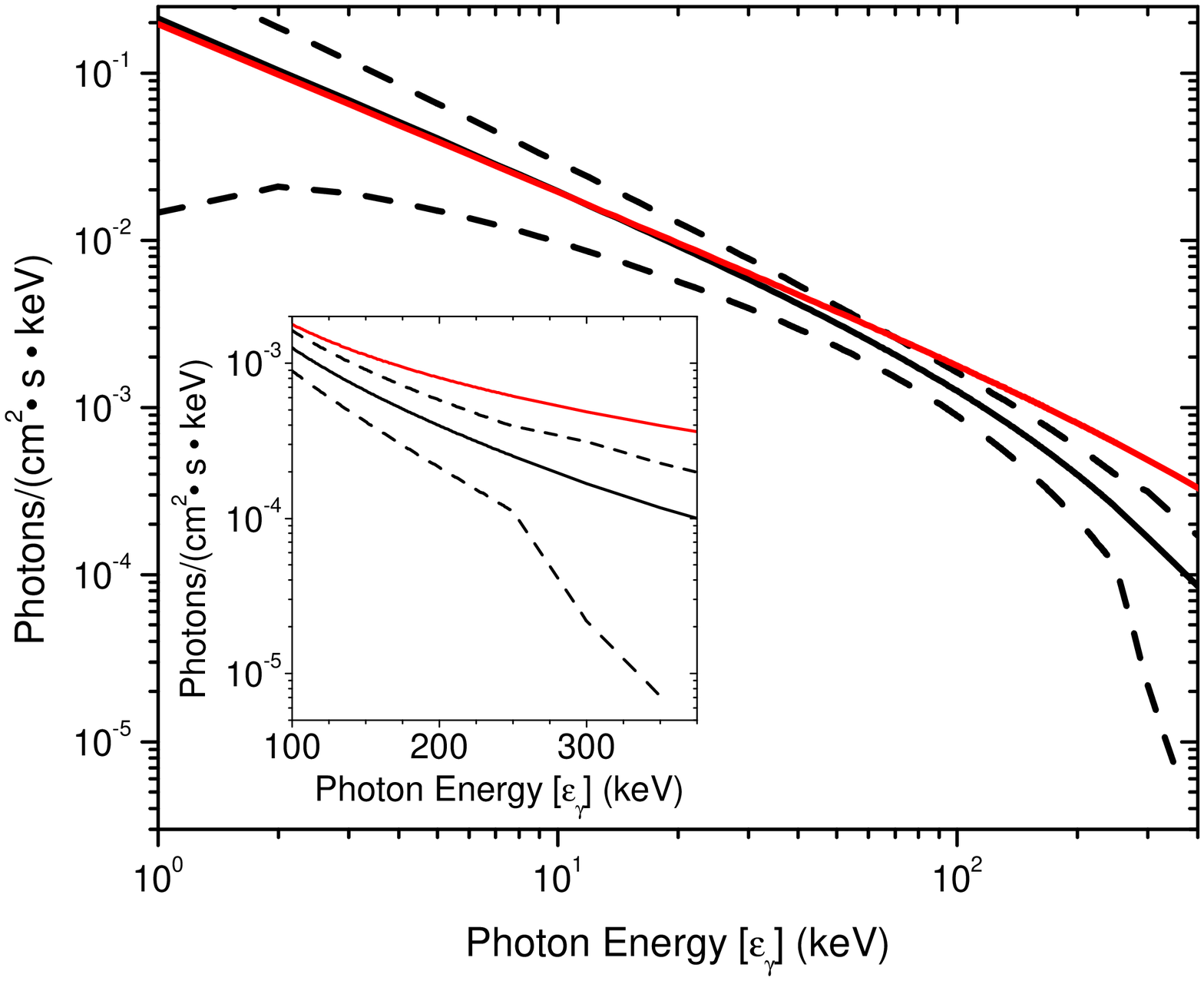}
\end{minipage}
\begin{minipage}[c]{0.46\textwidth}
\includegraphics[width=1.08\textwidth]{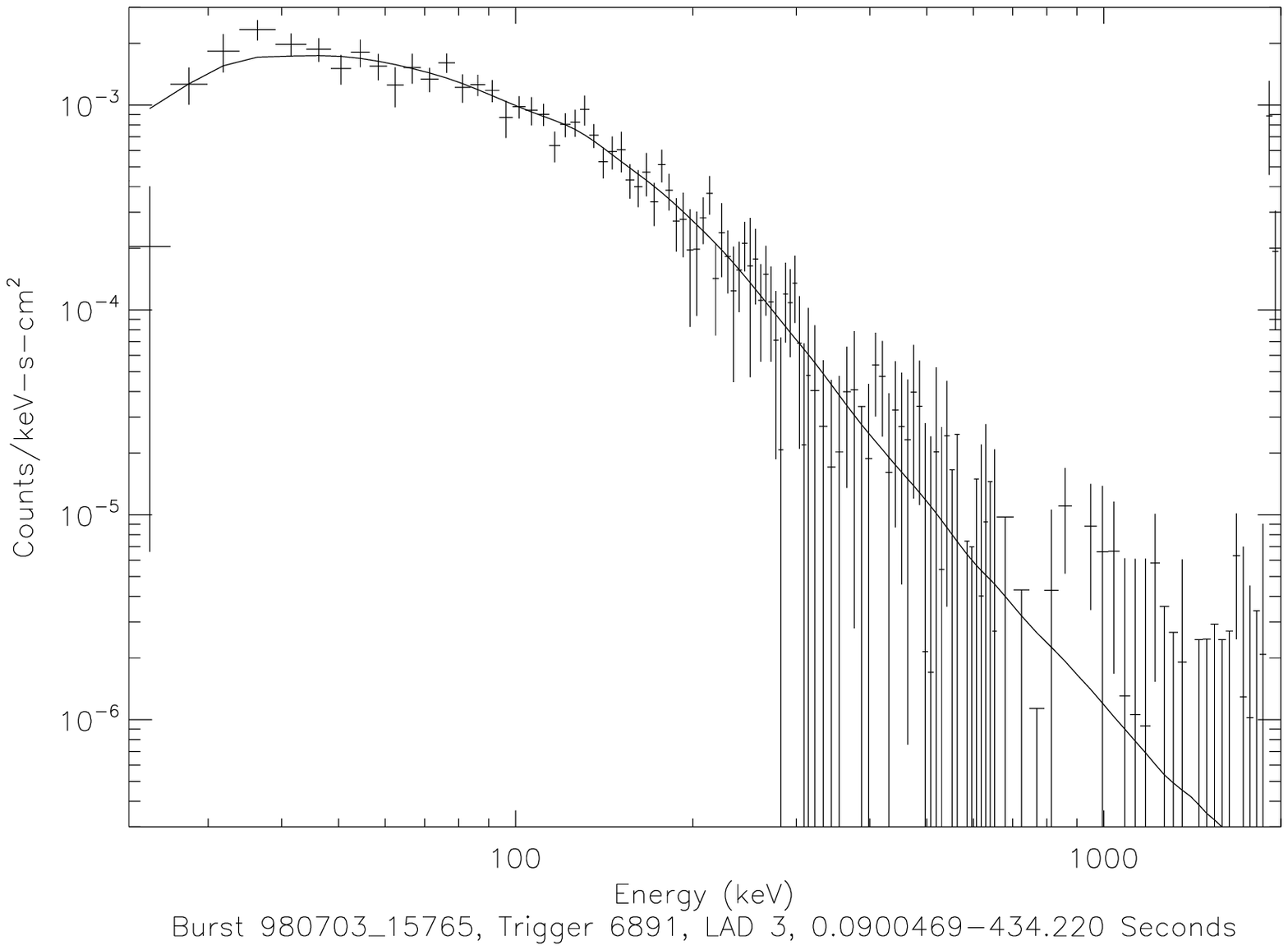}
\end{minipage}
\begin{minipage}[t]{0.25\textwidth}
\label{6891_fit_all}
\end{minipage}
\hspace{0.5cm}
\begin{minipage}[t]{0.25\textwidth}
\caption{\emph{\textbf{Left Plate}} - Prompt photon energy spectral fit to the Band function (see Equation~\ref{Band_function2}) for GRB980703a using the observed discrete electromagnetic parameters (solid black - used in models 1, 2, and 4), and average values (solid red - used in model 3). Inset illustrates the regime of the photon break energy ($\epsilon_{\gamma}^{b}$) for the discrete parameter fit. Note that the disagreement between the curves is not resolved by the uncertainty in the discrete fit, illustrated as 1$\sigma$ dashed confidence bands (assuming propagated errors with trivial covariance). See Tables \ref{GRB980703_observed _em_properties} and \ref{GRB980703_estimated_em_properties} for fit parameters. \emph{\textbf{Right Plate}} - The prompt photon energy spectral Band fit, convolved with BATSE detector response (solid line), is compared to the measured BATSE photon count rate (data points with error bars) for the entire (fluence) emission of GRB980703a.}
\label{Band_fit_GRB6891}
\end{minipage}
\end{figure}

Since the expected neutrino emission relies on the microphysics associated with the prompt $\gamma$-ray emission phase, a complete electromagnetic characterization of the GRB is compulsory. Spectral analysis, resulting in a prompt $\gamma$-ray photon energy spectrum, illustrated in Figure~\ref{Band_fit_GRB6891}, was performed via convolving an assumed spectral model with the BATSE detector response and comparing the fit with observed data. Our spectral assumption is characterized by an empirical model known as the \emph{Band function} \citep{Band:1993}:

\begin{equation}
N_{\epsilon_{\gamma}}(\epsilon_{\gamma})=
\begin{cases}
A_{\gamma}\left(\dfrac{\epsilon_{\gamma}}{100\:\text{keV}}\right)^{\alpha}exp\left(-\dfrac{\epsilon_{\gamma}}{\epsilon_{\gamma}^{o}}\right) & \epsilon_{\gamma}\leq\epsilon_{\gamma}^{b}\equiv(\alpha-\beta)\epsilon_{\gamma}^{o}\\
A_{\gamma}^{\prime}\left[\dfrac{(\alpha-\beta)\epsilon_{\gamma}^{o}}{100\:\text{keV}}\right]^{\alpha-\beta}exp(\beta-\alpha) \left(\dfrac{\epsilon_{\gamma}}{100\:\text{keV}}\right)^{\beta} & \epsilon_{\gamma}\geq\epsilon_{\gamma}^{b}\equiv(\alpha-\beta)\epsilon_{\gamma}^{o}
\end{cases}
\label{Band_function2}
\end{equation}

Multi-wavelength afterglow observations have resulted in a more comprehensive electromagnetic characterization of GRB980703a, such as an observed spectroscopic redshift, $z_{obs}$, via analysis of the host galaxy \cite{Djorgovski:1998}. Since the blast wave evolution is sensitive to intrinsic burst properties such as the explosion energy, emission geometry and circumstellar density, $n$, the afterglow has been used to deduce anisotropic emission from a jet break time, $t_{jet}$, and the relative energy fractions imparted to both electrons, $\epsilon_{e}$, and the magnetic field, $\epsilon_{B}$. The best fit solution is consistent with the canonical fireball model, with a jet emission geometry that is inferred by assuming a relativistic outflow, beamed into an ambient medium of constant density, within a 
collimated jet of half angle, $\theta_{\text{jet}}$ \cite{Frail:2003}. A beaming fraction correction, $f_{B}$, is then used to correct isotropic values for luminosity, $L_{\gamma}^{\text{iso}}$, and energy, $E_{\gamma}^{\text{iso}}$, into beam-corrected (jet) values of $L_{\gamma}^{\text{jet}}$ and $E_{\gamma}^{\text{jet}}$, respectively. Electromagnetic observables are summarized in Table~\ref{GRB980703_observed _em_properties}, while the electromagnetic parameterization of each model is given in Table~\ref{GRB980703_estimated_em_properties}.

\begin{table} [h]
\begin{tabular}{lcc}
\hline
\small{\textbf{Parameter(s)}} & \small{\textbf{Value}} & \small{\textbf{Reference}}\\
\hline
\footnotesize{RA, DEC, $\sigma_{\text{R}}\:(^{\circ})$ [J2000]} & \footnotesize{359.777775417, 8.585303861, 1.3$\overline{8}\times10^{-7}$} & \cite{Taylor:1998} \\
\footnotesize{T (UTC$^{\text{s}}$), T$_{90}$ (s)\footnote{The start of the $T_{90}$ interval, with respect to the trigger time ($T$), was -6.14 seconds \cite{BATSE:2003}.} [50-350 keV]} & \footnotesize{15765.22, 411.65 $\pm$ 9.27} & \cite{BATSE:2003} \\
\footnotesize{$F_{\gamma}^{\text{Total}}$ (ergs/cm$^{2}$) [50-350 keV]} &  \footnotesize{$(6.22\pm0.43)\times10^{-5}$} & \cite{BATSE:2003} \\
\footnotesize{$\Phi_{\gamma}^{\text{Peak}}$ (ergs/cm$^{2}$/s) [20-1000 keV]\tablenote{We take the energy band pass of 20-1000 keV as bolometric.}} & ($1.56\pm0.07)\times10^{-6}$ & \cite{Stamatikos:2006} \\
\footnotesize{$\alpha$, $\beta$, $\epsilon_{\gamma}^{o}$ (keV) [50-350 keV]\tablenote{$A_{\gamma}=(1.97\pm0.43)\times10^{-3}$ photons/cm$^{2}$/keV/s,  $\chi^{2}_{\nu}\sim0.80$, and signal/noise $\sim$ 4.11 (see Equation~\ref{Band_function2}).}} &  \footnotesize{$-1.02\pm0.20$, $-2.33\pm0.57$, $223\pm96$} & \cite{Stamatikos:2006} \\
\footnotesize{$\epsilon_{\gamma}^{b}$, $\epsilon_{\gamma}^{p}$ (keV) [50-350 keV]} &  \footnotesize{$293\pm185$, $219\pm104$} & \cite{Stamatikos:2006} \\
\footnotesize{$\overline{z}_{obs}$\tablenote{Based upon an average from emission and absorption (Doppler redshift) lines from the host galaxy \cite{Djorgovski:1998}.}} & \footnotesize{$0.9660\pm0.0006$} & \cite{Djorgovski:1998} \\
\footnotesize{$d_{L}$\tablenote{$\Lambda_{\text{CDM}}$ cosmology: H$_{\text{o}}$ = 72 $\pm$ 5 km/Mpc/s, $\Omega_{\text{m}}$ = 0.29 $\pm$ 0.07, $\Omega_{\Lambda}$ = 0.73 $\pm$ 0.07 \cite{Spergel:2003}, is utilized throughout.}} & $6219\pm2843$ Mpc $\sim(1.92\pm0.88)\times10^{28}$ cm & \cite{Stamatikos:2006} \\
\footnotesize{$\epsilon_{e},\:\epsilon_{B}$} & \footnotesize{$0.27\pm0.03$, $0.0018_{-0.0003}^{+0.0004}$} & \cite{Yost:2003} \\
\footnotesize{t$_{\text{jet}}$ (days)} & 3.43$\pm$0.50 & \cite{Frail:2003} \\
\footnotesize{n (cm$^{-3}$)} & 28$\pm$10 & \cite{Frail:2003} \\
\footnotesize{$\theta_{\text{jet}}$\footnote{$\theta_{\text{jet}}\approx0.101\:\text{rad}\left(\frac{t_{\text{jet}}}{1\:\text{day}}\right)^{\frac{1}{8}} \left(\frac{\xi}{0.2}\right)^{\frac{1}{8}} \left(\frac{n}{10\;\text{cm}^{-3}}\right)^{\frac{1}{8}} \left[\frac{(1+z)}{2}\right]^{-\frac{3}{8}} \left(\frac{E_{\gamma}^{\text{iso}}}{10^{53}\;\text{ergs}}\right)^{-\frac{1}{8}}$, with $\xi\sim0.20_{-0.20}^{+0.80}$ (see \cite{Friedman:2005}).}} & \footnotesize{$\sim(10.10\pm1.36)^{\circ}\approx(0.176\pm0.024$) rad} & \cite{Stamatikos:2006} \\
\footnotesize{$f_{B}\equiv1-\cos\theta_{\text{jet}}$} & $0.016\pm0.004$ & \cite{Stamatikos:2006} \\
\hline
\end{tabular}
\caption{Observed electromagnetic properties of GRB980703a (BATSE Trigger \#: 6891).}
\label{GRB980703_observed _em_properties}
\end{table}

\begin{table} [h]
\begin{tabular}{lcccc}
\hline
\small{\textbf{Parameter}} & \small{\textbf{Model 1\tablenote{Based upon discrete electromagnetic parameters, assuming isotropic emission.}}} & \small{\textbf{Model 2\tablenote{Based upon discrete electromagnetic parameters, assuming beamed (jet) emission.}}} & \small{\textbf{Model 3\tablenote{Using averaged GRB values: $\alpha\sim-1$, $\beta\sim-2$, $\epsilon_{\gamma}^{o}\sim\epsilon_{\gamma}^{b}\sim\epsilon_{\gamma}^{p}\sim1$ MeV, $F_{\gamma}^{\text{Total}}\sim6\times10^{-6}$ ergs/cm$^{2}$, $z\sim1$, and $\Phi_{\gamma}^{\text{Peak}}\sim2\times10^{-6}$ ergs/cm$^{2}$/s \cite{WaxmanBahcall:1999}. Under the assumption of isotropic emission.}}} & \small{\textbf{Model 4\tablenote{Isotropic emission and estimated $z$ via the lag-luminosity method, with $\tau_{o}=0.39_{-0.20}^{+0.14}$ seconds \citep{Bandref:2004}.}}} \\
\hline
\smallskip
\footnotesize{z} & \footnotesize{\emph{See Table~\ref{GRB980703_observed _em_properties}}} & \footnotesize{\emph{See Table~\ref{GRB980703_observed _em_properties}}} & \footnotesize{$\sim1$} & \footnotesize{$0.6378_{-0.1398}^{+0.2222}$} \\
\smallskip
\footnotesize{$d_{L}$ (Mpc)} & \footnotesize{\emph{See Table~\ref{GRB980703_observed _em_properties}}} & \footnotesize{\emph{See Table~\ref{GRB980703_observed _em_properties}}} & \footnotesize{$6491\pm2887$} & \footnotesize{$3730_{-2522}^{+2688}$} \\
\smallskip
\footnotesize{L$_{\gamma}$} (10$^{51}$ ergs/s) [20-1000 keV] & \footnotesize{$7.21\pm6.60$} & \footnotesize{$0.11\pm0.11$}  & \footnotesize{$10.08\pm8.97$} & \footnotesize{$2.38_{-1.20}^{+2.29}$} \\
\smallskip
\footnotesize{E$_{\gamma}$} (10$^{52}$ ergs) [20-2000 keV]  & \footnotesize{$13.77\pm12.68$} & \footnotesize{$0.21\pm0.21$}  & \footnotesize{$1.51\pm1.35$}  & \footnotesize{$6.57_{-2.67}^{+5.26}$} \\
\smallskip
\footnotesize{$\epsilon_{e}$} & \footnotesize{\emph{See Table~\ref{GRB980703_observed _em_properties}}} & \footnotesize{\emph{See Table~\ref{GRB980703_observed _em_properties}}} & \footnotesize{$0.33_{-0.33}^{+0.67}$} & \footnotesize{$0.33_{-0.33}^{+0.67}$} \\
\smallskip
\footnotesize{$\epsilon_{B}$} & \footnotesize{\emph{See Table~\ref{GRB980703_observed _em_properties}}} & \footnotesize{\emph{See Table~\ref{GRB980703_observed _em_properties}}} & \footnotesize{$0.33_{-0.33}^{+0.67}$} & \footnotesize{$0.33_{-0.33}^{+0.67}$} \\
\hline
\end{tabular}
\caption{Electromagnetic parameterization for GRB980703a neutrino models.}
\label{GRB980703_estimated_em_properties}
\end{table}

\begin{figure}[t]
\centering
\begin{minipage}[c]{0.46\textwidth}
\includegraphics[width=1.10\textwidth]{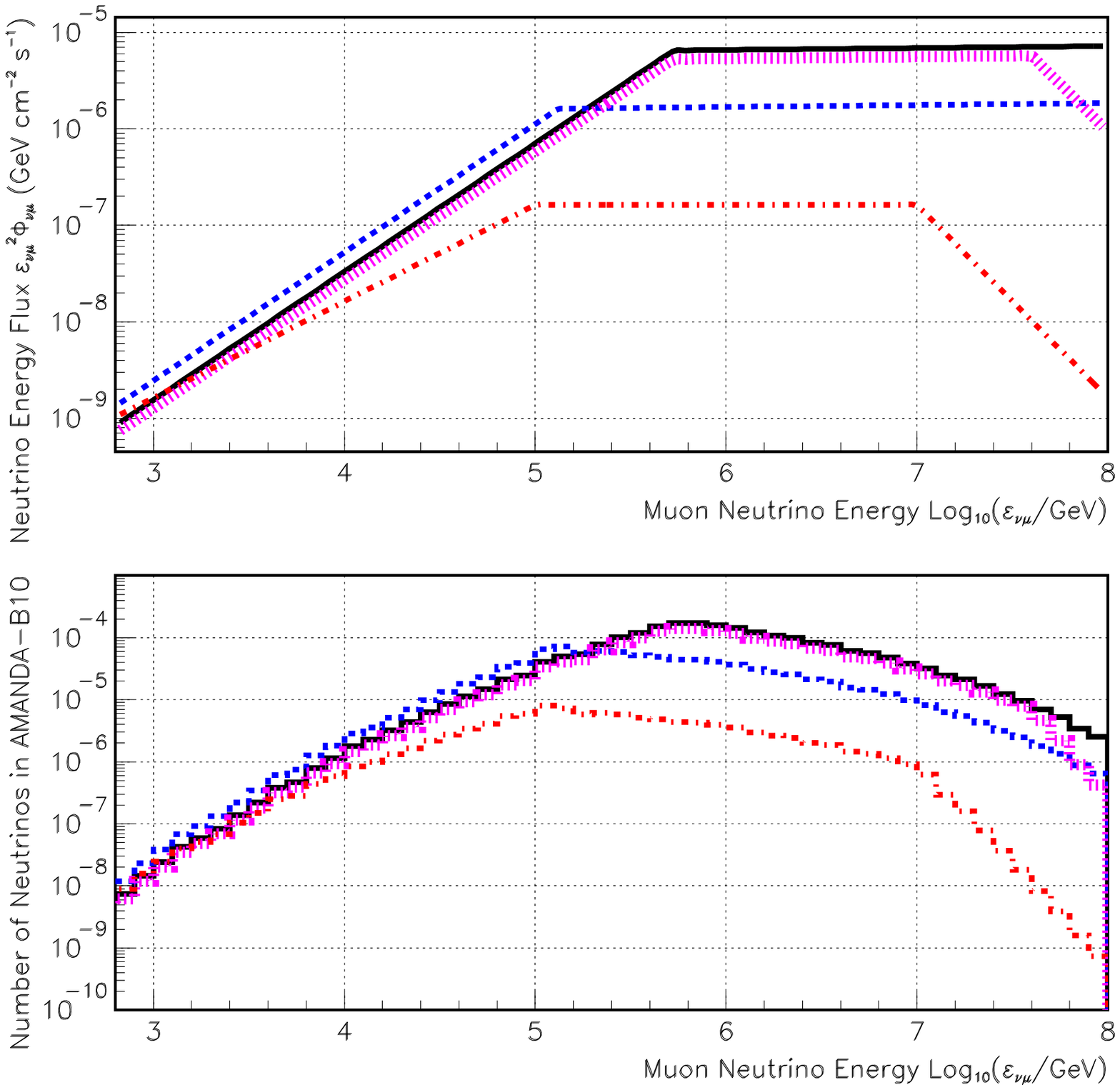}
\end{minipage}
\begin{minipage}[c]{0.46\textwidth}
\includegraphics[width=1.04\textwidth]{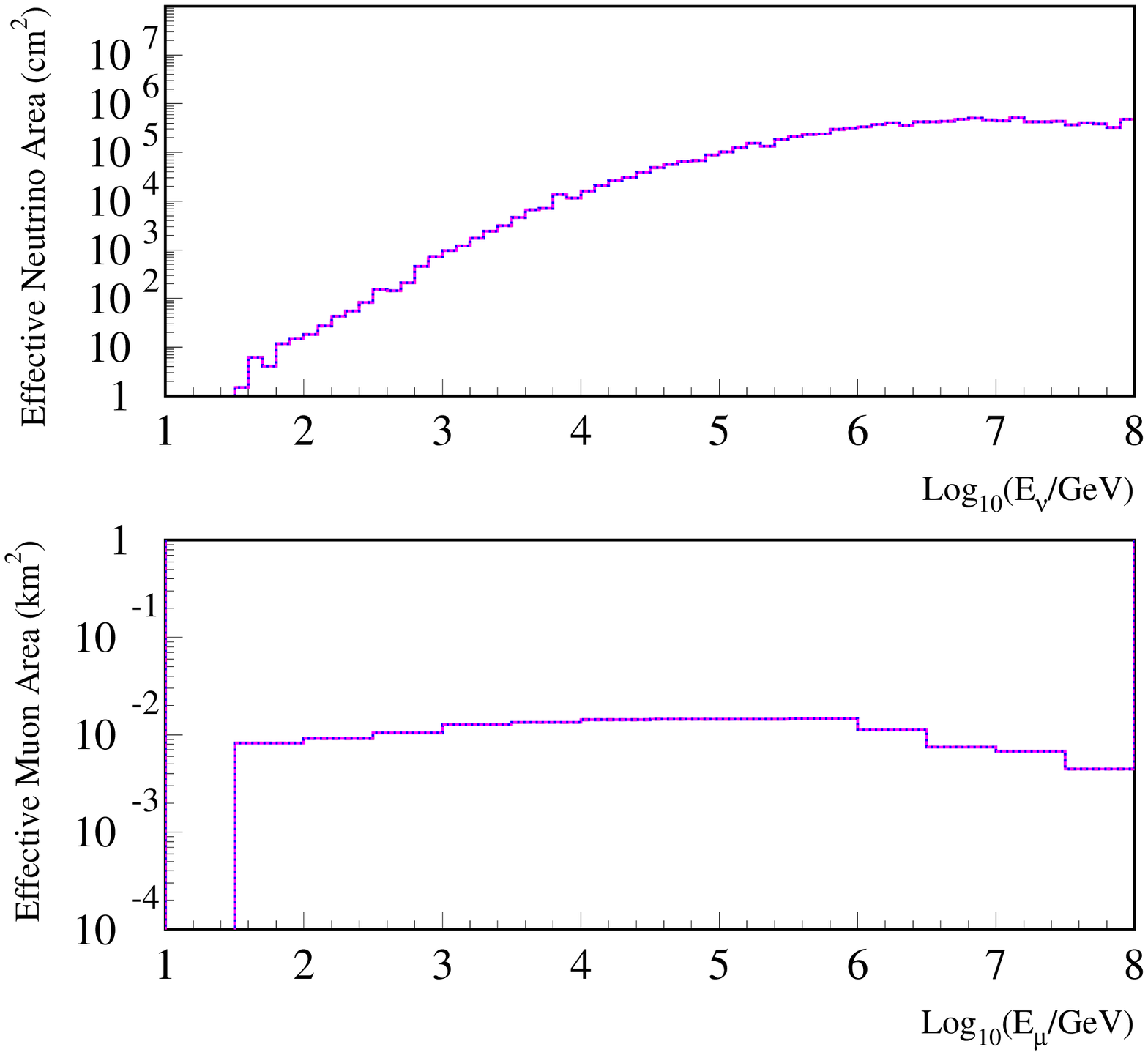}
\end{minipage}
\begin{minipage}[t]{0.25\textwidth}
\label{neutrino_spectrum_and_response2}
\end{minipage}
\hspace{0.5cm}
\begin{minipage}[t]{0.25\textwidth}
\caption{\emph{\textbf{Left Plate}} - Prompt neutrino energy flux (\emph{Upper panel}) and AMANDA-B10 detector response (\emph{Lower panel}), for GRB980703a models 1, 2, 3, and 4, indicated by solid black, dashed blue, dot-dashed red, and hatched magenta curves, respectively. \emph{\textbf{Right Plate}} - Effective area of AMANDA-B10 for muon neutrinos (\emph{Upper panel}) and muons (\emph{Lower panel} - as a function of energy at closest approach to the detector). Color code is identical to left plate. See Table~\ref{GRB980703_neutrino_parameters} for selection and signal efficiency details.}
\label{GRB980703a_detector_response}
\end{minipage}
\end{figure}

The neutrino spectral parameterization is given in Equation~\ref{neutrino_energy_spectrum} \cite{Stamatikos:2005d}. Values from Tables~\ref{GRB980703_observed _em_properties} and \ref{GRB980703_estimated_em_properties} have been substituted into Equation~\ref{neutrino_energy_spectrum} to produce the values of the neutrino parameters and the number of neutrino events expected in AMANDA-B10 (via simulation) for models 1-4, given in Table~\ref{GRB980703_neutrino_parameters}. The response of AMANDA-B10 to GRB980703a models 1-4, including effective areas for neutrinos and muons, is illustrated in Figure~\ref{GRB980703a_detector_response}.

\begin{equation}
\epsilon_{\nu_{\mu}}^{2}\Phi_{\nu_{\mu}}\approx A_{\nu_{\mu}}\times
\begin{cases}
\left(\frac{\epsilon_{\nu_{\mu}}}{\epsilon_{\nu}^{b}}\right)^{-\beta-1} & \epsilon_{\nu_{\mu}}<\epsilon_{\nu}^{b} \\
\left(\frac{\epsilon_{\nu_{\mu}}}{\epsilon_{\nu}^{b}}\right)^{-\alpha-1} & \epsilon_{\nu}^{b}<\epsilon_{\nu_{\mu}}<\epsilon_{\pi}^{b} \\
\left(\frac{\epsilon_{\nu_{\mu}}}{\epsilon_{\nu}^{b}}\right)^{-\alpha-1}\left(\frac{\epsilon_{\nu_{\mu}}} {\epsilon_{\pi}^{b}}\right)^{-2} & \epsilon_{\nu_{\mu}}>\epsilon_{\pi}^{b}
\end{cases}
\label{neutrino_energy_spectrum}
\end{equation}

\begin{table} [h]
\begin{tabular}{lcccc}
\hline
\small{\textbf{Parameter\tablenote{Where $L_{\gamma}\equiv L_{\gamma,52}\cdot10^{52}\:\text{ergs/s}$, $\Gamma\equiv\Gamma_{2.5}\cdot10^{2.5}$, $t_{v}\equiv t_{v,-2}\cdot 10\:\text{ms}$, $\epsilon_{\gamma}^{b}\equiv\epsilon_{\gamma,\text{MeV}}^{b}\cdot1\:\text{MeV}$, and $\epsilon_{\gamma}^{max}\equiv\epsilon_{\gamma,MeV}^{max}\cdot100\:\text{MeV}$.}}} & \small{\textbf{Model 1}} & \small{\textbf{Model 2}} & \small{\textbf{Model 3}\tablenote{Bracketed values indicate nominal values used in average GRB parameterization \cite{WaxmanBahcall:1999}.}} &  \small{\textbf{Model 4}} \\
\hline
\smallskip
\footnotesize{$\Gamma_{\text{bulk}}\gtrsim276\left[L_{\gamma,52}t_{v,-2}^{-1}\epsilon_{\gamma,MeV}^{max}(1+z)\right]^{\frac{1}{6}}$ \tablenote{Super-Eddington luminosity within a compact source requires a lower bound to ensure transparent optical depth.}} & \footnotesize{$293\pm45$} & \footnotesize{$146\pm23$} & \footnotesize{$310\pm46$ [300]} & \footnotesize{$243_{-17}^{+33}$} \\
\smallskip
\footnotesize{$f_{\text{$\pi$}}\simeq 0.2\times \frac{L_{\gamma,52}}{\Gamma_{2.5}^{4}t_{v,-2}\epsilon_{\gamma, MeV}^{b}(1+z)}$} & \footnotesize{$0.34\pm0.43$} & \footnotesize{$0.09\pm0.11$} & \footnotesize{$0.11\pm0.12$ [0.20]} & \footnotesize{$0.34_{-0.28}^{+0.40}$} \\
\smallskip
\footnotesize{$A_{\nu_{\mu}}\approx\frac{F_{\gamma} f_{\pi}}{8\epsilon_{e}\ln(10)T_{90}}$ ($10^{-6}$ GeV/cm$^{2}$/s)} & \footnotesize{$6.49\pm8.27$} & \footnotesize{$1.62\pm2.13$} & \footnotesize{$0.16_{-0.24}^{+0.37}$} & \footnotesize{$5.27_{-6.81}^{+12.36}$} \\
\smallskip
\footnotesize{$\epsilon_{\nu}^{b}\approx\dfrac{7\times10^{5}}{(1+z)^{2}} \dfrac{\Gamma_{2.5}^{2}}{\epsilon_{\gamma,\text{MeV}}^{b}}$ ($10^{5}$ GeV)} & \footnotesize{$5.29\pm3.71$} & \footnotesize{$1.32\pm0.93$} & \footnotesize{$1.68\pm0.50$ [1]} & \footnotesize{$5.25_{-3.51}^{+3.88}$} \\
\smallskip
\footnotesize{$\epsilon_{\pi}^{b}\approx\frac{10^{8}\sqrt{\epsilon_{e}}\:\Gamma_{2.5}^{4}t_{v,-2}}{(1+z)\sqrt{\epsilon_{B}}\sqrt{L_{\gamma,52}}}$ ($10^{8}$ GeV)} & \footnotesize{$5.37_{-4.13}^{+4.15}$} & \footnotesize{$2.68_{-2.15}^{+2.16}$} & \footnotesize{$0.46_{-0.47}^{+0.75}$ [0.1]} & \footnotesize{$0.40_{-0.32}^{+0.64}$} \\
\smallskip
\footnotesize{$n_{\nu_{\mu}}^{\text{AMANDA-B10}}$ $(10^{-4})$\tablenote{Events expected in AMANDA-B10, on a background of $\sim$8.17, for an on-time search window of 420.92 seconds within a search bin radius of $\sim9.5^{\circ}$. Estimated event selection attenuates signal efficiency by $\sim75\%$, while rejecting $\sim99.7\%$ of the background. These values are preliminary estimates, a full analysis is currently in progress.}} & \footnotesize{6.08} & \footnotesize{2.79} & \footnotesize{0.32} & \footnotesize{4.95} \\
\hline
\end{tabular}
\caption{Prompt muon neutrino flux parameterization and event number for GRB980703a.}
\label{GRB980703_neutrino_parameters}
\end{table}

\section{Discussion \& Future Outlook in the Swift Era}

Consistent with our results for GRB030329, we find that the most critical parameters, which translate into an observable variation in detector response, are the electromagnetic fluence, $F_{\gamma}$, and spectral characterization in the vicinity of the photon break energy, $\epsilon_{\gamma}^{b}$. The former is related to the number of neutrinos expected in the detector, while the latter affects the mean neutrino energy of the events. These effects on neutrino energy flux, which exceed one and two orders of magnitude in the comparison of models 1 and 3, in GRB980703a and GRB030329, respectively, may be directly traced back to the variance in their electromagnetic characterization, which directly affects the constraints placed upon astrophysical models in the case of null detection. The consequences of this reality are both unequivocal and apropos, since a hallmark of the Swift era is the acquisition of a more complete electromagnetic characterization of fewer bursts (relative to the age of BATSE). With an order of magnitude increase in effective area for GRBs \cite{Stamatikos:2005d}, realized by km-scale detectors such as IceCube, it is most likely that either the first evidence for correlated neutrino emission or the first real constraints on associated hadronic acceleration will come from the analysis of an exceptional (local) discrete GRB, rather than an aggregate of hundreds with average emission. It is interesting to note that model 4, based upon an estimated redshift from the lag-luminosity relation, was consistent with model 1, which was based upon the observed redshift. Future work includes the analysis of a subset of BATSE GRBs from 1997-2000 \cite{Stamatikos:2004b,Stamatikos:2006}, using Band function fits and (lag-luminosity relation) estimated redshifts. Ultimately, a synergy of gamma-ray and neutrino astronomy may be realized, within the context of GRBs, via multi-wavelength and multi-messenger correlative observational campaigns in an era of superior scientific instruments such as IceCube and Swift.


\bibliographystyle{aipproc}   

\bibliography{Stamatikos_Swift_Symposium_Proceedings_References}

\hyphenation{Post-Script Sprin-ger}
\begin{thebibliography}{16}
\expandafter\ifx\csname natexlab\endcsname\relax\def\natexlab#1{#1}\fi
\providecommand{\enquote}[1]{``#1''}
\expandafter\ifx\csname url\endcsname\relax
  \def\url#1{\texttt{#1}}\fi
\expandafter\ifx\csname urlprefix\endcsname\relax\def\urlprefix{URL }\fi
\providecommand{\eprint}[2][]{\url{#2}}

\bibitem[Waxman and Bahcall(1999)]{WaxmanBahcall:1999}
E.~Waxman, and J.~Bahcall, \emph{Phys. Rev. D} \textbf{59}, 023002 (1999).

\bibitem[Achterberg et~al.(See also K. Kuehn et al., these
  proceedings.)]{Achterberg:2005}
A.~Achterberg, et~al., \emph{astro-ph/0509330}  (See also K. Kuehn et al.,
  these proceedings.).

\bibitem[Guetta et~al.(2004)]{Guettaref:2004}
D.~Guetta, et~al., \emph{Astroparticle Physics} \textbf{20}, 429--455 (2004).

\bibitem[{Stamatikos} et~al.(2005)]{Stamatikos:2005d}
M.~{Stamatikos}, J.~{Kurtzweil}, and M.~{Clarke}, \emph{preprint
  (astro-ph/0510336)}  (2005).

\bibitem[Stamatikos et~al.(2004)]{Stamatikos:2004b}
M.~Stamatikos, et~al., \enquote{Gamma-Ray Bursts: 30 Years of Discovery,} AIPC
  727, 2004, pp. 146--149.

\bibitem[{Stamatikos} et~al.(2006)]{Stamatikos:2006}
M.~{Stamatikos}, D.~{Band}, D.~{Hooper}, and F.~{Halzen}, \emph{In preparation,
  to be submitted to ApJ.}  (2006).

\bibitem[Band et~al.(2004)]{Bandref:2004}
D.~Band, et~al., \emph{ApJ} \textbf{613}, 484--491 (2004).

\bibitem[{Hill} et~al.(2005)]{Hill:2005}
G.~{Hill}, J.~{Hodges}, B.~{Hughey}, A.~{Karle}, and M.~{Stamatikos},
  \enquote{PHYSTAT,} Oxford, 2005.

\bibitem[Band et~al.(1993)]{Band:1993}
D.~Band, et~al., \emph{ApJ} \textbf{413}, 281--292 (1993).

\bibitem[Djorgovski et~al.(1998)]{Djorgovski:1998}
S.~Djorgovski, et~al., \emph{ApJ} \textbf{508}, L17--L20 (1998).

\bibitem[Frail et~al.(2003)]{Frail:2003}
D.~Frail, et~al., \emph{ApJ} \textbf{590}, 992 (2003).

\bibitem[Taylor et~al.(1998)]{Taylor:1998}
G.~Taylor, et~al., \emph{GCN GRB Observation Report 152}  (1998).

\bibitem[BAT(2003)]{BATSE:2003}
  (2003),
  \urlprefix\url{http://www.batse.msfc.nasa.gov/batse/grb/catalog/current/}.

\bibitem[Yost et~al.(2003)]{Yost:2003}
S.~Yost, et~al., \emph{ApJ} \textbf{597}, 459--473 (2003).

\bibitem[Friedman and Bloom(2005)]{Friedman:2005}
A.~Friedman, and J.~Bloom, \emph{ApJ} \textbf{627}, 1--25 (2005).

\bibitem[Spergel et~al.(2003)]{Spergel:2003}
D.~Spergel, et~al., \emph{ApJ Supp. Ser.} \textbf{148}, 174--194 (2003).

\end{thebibliography}

\IfFileExists{\jobname.bbl}{}
 {\typeout{}
  \typeout{******************************************}
  \typeout{** Please run "bibtex \jobname" to optain}
  \typeout{** the bibliography and then re-run LaTeX}
  \typeout{** twice to fix the references!}
  \typeout{******************************************}
  \typeout{}
 }

\end{document}